\documentclass[aps,pre,floatfix]{revtex4}

\usepackage{graphicx}
\usepackage{amssymb,amsfonts,amsmath}
\usepackage{epsf}
\usepackage{subfigure}
\usepackage{epstopdf}
\usepackage{mathrsfs}

\font\helvb=cmssbx12

\newcommand{\be}{\begin{equation}}
\newcommand{\ee}{\end{equation}}
\newcommand{\bea}{\begin{eqnarray}}
\newcommand{\eea}{\end{eqnarray}}

\begin{document}

\title{\bf Trace formula for activated escape in noisy maps}

\author{J. Demaeyer and P. Gaspard}
\affiliation{Center for Nonlinear Phenomena and Complex Systems, Universit\'e Libre de Bruxelles, Code Postal 231, Campus Plaine, B-1050 Brussels, Belgium}

\begin{abstract}
Using path-integral methods, a formula is deduced for the noise-induced escape rate from an attracting
fixed point across an unstable fixed point in one-dimensional maps.  The calculation starts from the trace formula for the eigenvalues of the Frobenius-Perron operator ruling the time evolution of the
probability density in noisy maps.  The escape rate is determined from the loop formed by two
heteroclinic orbits connecting back and forth the two fixed points of the one-dimensional map
extended to a two-dimensional symplectic map.  The escape rate is obtained with the expression of
the prefactor to Arrhenius-van't Hoff exponential factor.
\end{abstract}

\noindent 

\vskip 0.5 cm

\maketitle

\section{Introduction}

The escape of trajectories is a ubiquitous phenomenon in open dynamical systems and stochastic processes.  By this basic mechanism, trajectories may leave without return a bounded phase-space region or a metastable state.  If escape occurs repetitively for a statistical ensemble of trajectories, the population of remaining trajectories often undergoes an exponential decay characterized by the so-called escape rate.  Its inverse defines the lifetime of the decaying state, which represents an intrinsic property of the system.  This paradigm is fundamental to nucleation theory and reaction-rate theory in chemistry, physics, and biology \cite{K40,C43,vK81,G04,HTB90,ND12}.

In many circumstances, escape is activated by the presence of noise, which may be of internal or external origin.  This is the case for thermally activated escape over a potential energy barrier in Kramers problem \cite{K40} and, more generally, for noise-induced escape in continuous-time or discrete-time dynamics.  This latter concerns dissipative dynamical systems that are periodically driven or subjected to some cybernetic feedback \cite{GT84,GT85,SDRVBB99,NVPAVDGM04,N10}.  Discrete-time dynamics is also of application in oscillatory regimes dominated by a sufficiently well-defined period so that the continuous-time dynamics can be modeled with a Poincar\'e first-return map.  In the presence of noise, such systems can be described by noisy maps, which have been the topic of several studies \cite{RT91,RMT94,RT95a,RT95b,D99,CSPVD99,PVV01,BMLSMcC05,SBLMcC05,FJ05,DG09,RGM10}.

At the deterministic level of description, such dynamics may present an attractor surrounded by a basin of attraction.  In the presence of noise, leakage would occur at the border of the basin, inducing the escape of trajectories to infinity.  Typically, the rate of such activated escape processes vanishes with the noise amplitude in a non-analytic way, which is well known in chemical kinetics as the Arrhenius-van't Hoff law \cite{HTB90}. This non-analytic dependence expresses the fact that escape does not preexist in the corresponding deterministic system and is a novel phenomenon entirely generated by the noise.

The purpose of the present paper is to deduce a mathematical formula for the rate of activated escape in discrete-time dynamics.  With this aim, the escape rate is identified as the leading eigenvalue of the Frobenius-Perron operator ruling the time evolution of probability densities in the process \cite{D99,CSPVD99,PVV01}.  The spectrum of eigenvalues is given by a trace formula that is calculated with path-integral methods in the weak-noise limit.  In this limit, the path integral selects the classical orbits of a symplectic map, which is defined in a phase space extended to include momenta canonically conjugated to the variables of the deterministic map \cite{FJ05,DG09}.  Taking the trace of the iterates of the Frobenius-Perron operator is known to select closed orbits \cite{KT84,CE91,G98,D99,CSPVD99,PVV01}.  Here, a challenge arises because the closed orbits are linked with the fixed points corresponding, on the one hand, to the attractor and, on the other hand, to the top of the barrier over which escape is activated by the noise.  The closed orbits are thus forming a pair of heteroclinic orbits asymptotic to both fixed points.  For continuous-time dynamical systems, such heteroclinic orbits are called instantons or kinks and their effect has been much studied in the literature \cite{CCR81,S81}.  Here, our purpose is to consider discrete-time dynamics, for which we obtain the escape rate including the expression of the prefactor to Arrhenius-van't Hoff exponential factor.

The paper is organized as follows.  In Section~\ref{Trace}, we explain how the escape rate can be obtained from a trace formula for the Frobenius-Perron operator ruling the stochastic time evolution.  The weak-noise limit is considered in Section~\ref{Symplectic} where the symplectic approach is presented.  In Section~\ref{Calcul}, the trace is evaluated in the weak-noise limit around the pair of heteroclinic orbits linking both fixed points.  The evaluation of the path integral along the two heteroclinic orbits is carried out in Sections~\ref{ho1} and~\ref{ho2}.  The calculation is completed in Section~\ref{Rate} where the expression is finally obtained for the escape rate.  Section~\ref{Examples} presents the application of this result to the noisy logistic and exponential maps.  Conclusions are drawn in Section~\ref{Conclusions}.

\section{Trace formula for the Frobenius-Perron operator}
\label{Trace}

\subsection{Noisy maps}

We consider discrete-time dynamics in the one-dimensional space of the variable $x_n\in{\mathbb R}$.  The index $n\in{\mathbb N}$ is the discrete time.  The variable $x_n$ evolves in time according to
\be
x_{n+1}= f(x_n) + \xi_n
\label{noisy_map}
\ee
where $f(x)$ is a real function of $x\in{\mathbb R}$ and $\xi_n$ a sequence of independent Gaussian random variables characterized by 
\be
\langle\xi_n\rangle=0 \qquad\mbox{and}\qquad \langle\xi_n\xi_m\rangle=\epsilon \, \delta_{nm}
\ee
where $\epsilon$ is the parameter controlling the noise amplitude.

If the noise amplitude vanishes, $\epsilon=0$, the map becomes deterministic $x_{n+1}=f(x_n)$.  We suppose that this map has an attracting fixed point at $x=a$ with a basin of attraction extending from infinity to an unstable fixed point at $x=b$.  If $f'(x)=df/dx$ denotes the derivative of the function $f(x)$ with respect to its variable $x$, the linearized dynamics around the attractor is characterized by the factor $\Lambda_a=f'(a)$ satisfying $0<\Lambda_a<1$, so that a small perturbation $\delta x_n=x_n-a$ would evolve according to $\delta x_{n+1}=\Lambda_a \, \delta x_n$.  Around the unstable fixed point, the linearized dynamics is given by $\delta x_{n+1}=\Lambda_b \, \delta x_n$ with $\delta x_n=x_n-b$ and the stretching factor $\Lambda_b=f'(b)$ such that $\Lambda_b >1$.

\subsection{The Frobenius-Perron operator}

The probability density $\rho_n(x)$ that the trajectory is found in the position $x$ at the current time $n$ is ruled by the so-called Frobenius-Perron operator
\be
\rho_{n+1}(x) = \int_{-\infty}^{+\infty} dx_0 \, \rho_n(x_0) \, K(x_0,x) \equiv \hat P\rho_n(x) \; ,
\ee
which is defined in terms of the conditional probability density $K(x_0,x)$ to find the trajectory in the position $x$ if it was in the position $x_0$ at the previous iterate.  Since the noise is Gaussian of variance $\epsilon$, this conditional probability density is given by
\be
K(x_0,x) = \frac{1}{\sqrt{2\pi\epsilon}} \, {\rm e}^{-\frac{1}{2\epsilon} \left[ x-f(x_0)\right]^2} \, .
\label{kernel}
\ee
Accordingly, the probability density after $n$ iterates is expressed as $\rho_n=\hat P^n\rho_0$.
We notice that the Frobenius-Perron is not self-adjoint in general.

We consider the eigenvalue problem for the Frobenius-Perron operator
\be
\hat P \, \phi_\alpha(x) = \chi_\alpha \, \phi_\alpha(x)
\ee
where $\{\chi_\alpha\}$ are the eigenvalues and $\{\phi_\alpha(x)\}$ the associated eigenfunctions.  The  eigenfunctions of the adjoint operator $\hat P^{\dagger}$ are denoted $\{\tilde\phi_\alpha(x)\}$ and they satisfy the biorthonormality condition $\langle\tilde\phi_\alpha\vert\phi_\beta\rangle=\delta_{\alpha\beta}$ where $\langle \psi \vert \varphi\rangle=\int_{-\infty}^{+\infty} \psi^*(x)\, \varphi(x) \, dx$. If the eigenfunctions form a complete basis, the time evolution of the probability density could be decomposed as
\be
\rho_n(x) = \hat P^n\rho_0 = \sum_\alpha \langle\tilde\phi_\alpha\vert\rho_0\rangle \, \chi_\alpha^n \, \phi_\alpha(x) \; . 
\label{spectr_decomp}
\ee
Otherwise, the decomposition may have to include terms arising from Jordan-block structures or contributions from a continuous spectrum.  Since the probability density is always normalized to unity by the conservation of probability, the eigenvalues belong to the unit circle in the complex plane: $\vert\chi_\alpha\vert \leq 1$.  

Since the kernel (\ref{kernel}) of the operator is positive, the leading eigenvalue is simple and positive by the Frobenius-Perron theorem, and its associated eigenfunction is also positive \cite{G59}.  If the leading eigenvalue is equal to unity, the probability density converges towards a nonvanishing invariant density given by the eigenfunction $\phi_0(x)$ associated with the unit eigenvalue $\chi_0=1$.

In the case of escape, the leading eigenvalue is smaller than unity, $\chi_0<1$, and the probability density is vanishing in the long-time limit.  In this case, the escape rate can be given as $\gamma=-\ln\chi_0$ in terms of the leading eigenvalue.

\subsection{Fredholm determinant and trace formula}

The eigenvalues can be obtained as the zeros of the characteristic determinant also called the Fredholm determinant
\be
\det\left(\hat I - \chi^{-1}\, \hat P\right)=0
\ee
where $\hat I$ is the identity operator \cite{D99,CSPVD99,PVV01}.  Now, the logarithm of the determinant is equal to the trace of the logarithm and the logarithm of $\left(\hat I - \chi^{-1}\, \hat P\right)$ can be expanded in Taylor series to get
\be
\det\left(\hat I - \chi^{-1}\, \hat P\right)=\exp\left( - \sum_{n=1}^{\infty} \frac{1}{n\, \chi^n} \, {\rm tr}\, \hat P^n \right) \, .
\label{det-tr}
\ee
If the spectral decomposition (\ref{spectr_decomp}) holds, the trace of iterates of the Frobenius-Perron operator can be expressed in terms of the eigenvalues as
\be
{\rm tr}\, \hat P^n = \sum_\alpha \langle\tilde\phi_\alpha\vert\hat P^n \vert\phi_\alpha\rangle = \sum_\alpha \chi_\alpha^n \; .
\label{trace-eigen}
\ee
After substituting in Eq.~(\ref{det-tr}), we find that 
\be
\det\left(\hat I - \chi^{-1}\, \hat P\right)=\prod_\alpha \left( 1 - \frac{\chi_\alpha}{\chi}\right)=0 \; ,
\ee
which confirms that the eigenvalues are given by the zeros of the Fredholm determinant.

Alternatively, the expression (\ref{trace-eigen}) can also be used if the trace of iterates of the Frobenius-Perron operator may be calculated.  This latter can be written in terms of the corresponding kernel, which can be expressed as the following path integral
\bea
K_n(x_0,x) &=& \int dx_1 \, dx_2 \, \cdots \, dx_{n-1}\, K(x_0,x_1) \, K(x_1,x_2) \, \cdots \, K(x_{n-1},x) \nonumber\\
&=& \frac{1}{(2\pi\epsilon)^{n/2}} \, \int dx_1 \, dx_2 \, \cdots \, dx_{n-1}\, {\rm e}^{-\frac{1}{2\epsilon}\sum_{i=0}^{n-1}\left[ x_{i+1}-f(x_i)\right]^2}
\label{path-int}
\eea
with $x_n=x$.  The trace (\ref{trace-eigen}) is thus given by
\bea
{\rm tr}\, \hat P^n &=& \int dx \, K_n(x,x) \nonumber\\
&=& \frac{1}{(2\pi\epsilon)^{n/2}} \, \int dx_0 \, dx_1 \, dx_2 \, \cdots \, dx_{n-1}\, {\rm e}^{-\frac{1}{\epsilon}W_n(x_0,x_1,x_2,...,x_{n-1})}
\label{trace-int}
\eea
in terms of the action of the path
\be
W_n(x_0,x_1,x_2,...,x_{n-1})= \frac{1}{2} \sum_{i=0}^{n-1}\left[ x_{i+1}-f(x_i)\right]^2
\label{action}
\ee
with $x_n=x_0$.

In Appendix~\ref{AppA}, we show how this method applies to the simple case of noisy linear maps.

\section{Weak-noise limit and symplectic approach}
\label{Symplectic}

In the weak-noise limit, path integrals can be evaluated with the steepest-descent method, which selects the paths that are the extremals of the action functional
\be
\delta W_n=0 \; .
\ee
These extremals are the orbits of the second-order recurrence
\be
x_i-f(x_{i-1})-f'(x_i)\left[ x_{i+1}-f(x_i)\right]=0 \; ,
\label{2nd-order_map}
\ee
which is the coefficient of $\delta x_i$ in $\delta W_n$ with the notation $f'(x)=df(x)/dx$.  Introducing the momentum 
\be
p_i\equiv f'(x_i)\left[ x_{i+1}-f(x_i)\right] \; ,
\ee
which is canonically conjugated to the variable $x_i$, the second-order recurrence (\ref{2nd-order_map}) can be written in the form of the first-order symplectic map \cite{FJ05,DG09}
\be
\left\{
\begin{array}{l}
x_{i+1} = f(x_i) + \frac{p_i}{f'(x_i)} \, ,\\
p_{i+1} = \frac{p_i}{f'(x_i)} \, .
\end{array}
\right.
\label{sympl_map}
\ee

In the trace (\ref{trace-int}), these orbits are closed with $x_n=x_0$, which includes the fixed points and the genuine periodic orbits.

In the systems we consider, there exist two fixed points.  First, near the attractor at $(x=a,p=0)$, the symplectic map (\ref{sympl_map}) linearizes into
\be
\left(
\begin{array}{c}
\delta x_{i+1}\\
\delta p_{i+1}
\end{array}
\right)
=
\left(
\begin{array}{cc}
\Lambda_a & \Lambda_a^{-1}\\
0 & \Lambda_a^{-1}
\end{array}
\right)
\left(
\begin{array}{c}
\delta x_{i}\\
\delta p_{i}
\end{array}
\right)
\label{lin_map_a}
\ee
with $\Lambda_a=f'(a)$ for $\delta x_i=x_i-a$ and $\delta p_i=p_i$.  Since $\vert\Lambda_a\vert <1$, the  direction $\delta p =0$ is contracting by the factor $\Lambda_a$,  while the direction $\delta p=\left(1-\Lambda_a^2\right)\delta x$ is expanding by the factor $\Lambda_a^{-1}$.  The stable manifold $M_{\rm s}(a)$ of the attracting fixed point coincides with the axis $p=0$.  Instead, the unstable manifold $M_{\rm u}(a)$ extends to non-vanishing values of the momentum $p$.  

Secondly, near the fixed point $(x=b,p=0)$ at the barrier, the symplectic map linearizes into
\be
\left(
\begin{array}{c}
\delta x_{i+1}\\
\delta p_{i+1}
\end{array}
\right)
=
\left(
\begin{array}{cc}
\Lambda_b & \Lambda_b^{-1}\\
0 & \Lambda_b^{-1}
\end{array}
\right)
\left(
\begin{array}{c}
\delta x_{i}\\
\delta p_{i}
\end{array}
\right)
\label{lin_map_b}
\ee
with $\Lambda_b=f'(b)$ for $\delta x_i=x_i-b$ and $\delta p_i=p_i$.  Since $\vert\Lambda_b\vert >1$, the  direction $\delta p =0$ is expanding by the factor $\Lambda_b$,  while the direction $\delta p=\left(1-\Lambda_b^2\right)\delta x$ is contracting by the factor $\Lambda_b^{-1}$.  The unstable manifold $M_{\rm u}(b)$ of this unstable fixed point coincides with the axis $p=0$ and its stable manifold $M_{\rm s}(b)$ extends to non-vanishing values of the momentum~$p$.  We suppose that $b<a$.

The invariant manifolds of the fixed points are connected to each other.  

On the one hand, the interval $[b,a]$ extending between both fixed point on the axis $p=0$ constitutes the intersection between the stable manifold of the attracting fixed point and the unstable manifold of the unstable fixed point: $[b,a] = M_{\rm s}(a)\cap M_{\rm u}(b)$.  Therefore, this interval contains a continuum of heteroclinic orbits $h_{ba}$ going from the unstable fixed point towards the attracting one, $b\to a$, because
\be
\lim_{j\to + \infty} f^j(x) = a \qquad \mbox{and}\qquad \lim_{j\to -\infty} f^j(x) = b \qquad \forall \, x\in[b,a] \; .
\ee

On the other hand, the intersection of the unstable manifold of the attracting fixed point with the stable manifold of the unstable fixed point contains two heteroclinic orbits from the attracting fixed point to the unstable one $a\to b$: $M_{\rm u}(a)\cap M_{\rm s}(b)=h_{ab}^+\cup h_{ab}^-$.  These heteroclinic orbits differ by the orientation of the intersections of both invariant manifolds at every of their points.  

The invariant manifolds $M_{\rm u}(a)$ and $M_{\rm s}(b)$ form a typical heteroclinic structure.  As a consequence, the symplectic map is chaotic and there exist periodic orbits of arbitrarily large period $n$.  The periodic orbits forming a single loop between both fixed points fall in two families: the family $\{p_n^+\}$ accumulating to the heteroclinic orbit $h_{ab}^+$ and closing the loop in the vicinity of the heteroclinic orbits $h_{ba}$ and the other family $\{p_n^-\}$ accumulating to the heteroclinic orbit $h_{ab}^-$ and also closing the loop in the vicinity of the heteroclinic orbits $h_{ba}$.  Therefore, the loop is arbitrarily close to a pair of heteroclinic orbits: $h_{ab}^{\pm}\cup h_{ba}$.  In principle, the periodic orbits also contribute to the trace.  In the present systems, these periodic orbits come closer and closer to the fixed points and can thus be decomposed into a pair of heteroclinic orbits.

\section{The trace in the weak-noise limit}
\label{Calcul}

\subsection{The contribution of the fixed points and escape}

In the presence of the stable and unstable fixed points $a$ and $b$ connected by a pair of heteroclinic orbits, the noise induces escape over the barrier $b$.  Our purpose is thus to calculate the escape rate in the weak-noise limit by using the trace formula.

\begin{figure}[htbp]
\centerline{\includegraphics[width=7cm]{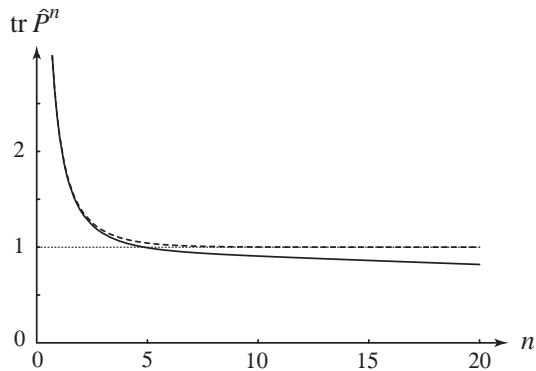}}
\caption{Typical time evolution of the trace of the Frobenius-Perron operator (solid line) compared with the approximation (\ref{trace-pts}) based on the two fixed points (dashed line).  The dotted line is the asymptote (\ref{trace-pts-asympt}) of this approximation.  The solid line shows the behavior (\ref{trace-exp}) expected including the escape over the barrier.}
\label{fig1}
\end{figure}

As aforementioned, the trace selects the periodic orbits including the two fixed points $a$ and $b$ \cite{G02}.  Each fixed point contributes by Eq.~(\ref{trace-pt}) obtained in Appendix~\ref{AppA}.  
Their contributions add to give
\be
{\rm tr}\, \hat P^n\Big\vert_{\rm fixed\, pts} =  \frac{1}{\vert 1-\Lambda_a^n\vert} + \frac{1}{\vert 1-\Lambda_b^n\vert} \; .
\label{trace-pts}
\ee
This function of the time $n$ is depicted as the dashed line in Fig.~\ref{fig1}.
Since $\vert\Lambda_a\vert <1$ for the attractor $a$ and $\vert\Lambda_b\vert >1$ for the unstable fixed point, the trace (\ref{trace-pts}) has the following long-time behavior 
\be
{\rm tr}\, \hat P^n\Big\vert_{\rm fixed\, pts} =  1 + O(\Lambda_a^n) + \frac{1}{\vert \Lambda_b\vert^n} + O\left(\frac{1}{\vert \Lambda_b\vert^{2n}}\right) \qquad\mbox{for}\qquad n\to\infty \; .
\label{trace-pts-asympt}
\ee
Therefore, the contribution of the fixed points approaches asymptotically the unit value, but does not decay exponentially as would be expected in the presence of escape. Instead, we expect by Eq. (\ref{trace-eigen}) with $\vert\chi_\alpha\vert <1$ that the trace should decay as
\be
{\rm tr}\, \hat P^n \simeq {\rm e}^{-\gamma \, n } \qquad\mbox{for}\qquad n\to\infty
\label{trace-exp}
\ee
where the escape rate $\gamma$ is given in terms of the leading eigenvalue as
\be
\gamma = -\ln\chi_0 \simeq A \, {\rm e}^{-W/\epsilon} \qquad\mbox{for}\qquad \epsilon\to 0
\label{Arrhenius}
\ee
with positive constants $A$ and $W$.  The expected behavior with this long-time exponential decay due to escape is shown as the solid line in Fig.~\ref{fig1}.  

The escape rate should thus have the non-analytic behavior in $\epsilon$ expected from Arrhenius-van't~Hoff law for activated processes.  In the weak-noise limit $\epsilon$, the trace (\ref{trace-exp}) can be further expanded as
\be
{\rm tr}\, \hat P^n =1 -\gamma \, n + O(\gamma^2 \, n^2) \; .
\label{trace-exp-Taylor}
\ee
Comparing with Eq.~(\ref{trace-pts-asympt}), the first term -- which is equal to unity -- is coming from the trace over the attracting fixed point, but extra orbits are expected to contribute to the next term $-\gamma \, n$ that should allow us to get the escape rate $\gamma$.  As shown in the following, these extra orbits are given by the pair of heteroclinic orbits connecting both fixed points $a$ and $b$.  Consequently, the trace has to be calculated by including the contribution of the loop formed by the pair of heteroclinic orbits $h_{ab}^{\pm}\cup h_{ba}$.

\subsection{The contribution of pairs of heteroclinic orbits}

The calculation of the path integral (\ref{trace-int}) for the contribution of the loop to the trace is organized as follows. The $n$ variables of integration $\{x_i\}_{i=0}^{n-1}$ are chosen in the vicinity of two orbits of the symplectic map (\ref{sympl_map}):

(1) The first belongs to the axis $p=0$, starting from an arbitrary point $x_0$ on this axis in the vicinity of the unstable fixed point $b$, and going to the attracting fixed point $a$ during $m-1$ iterations
\be
\bar{x}_i=\bar{x}_i(x_0)=f^{i}(x_0)  \qquad\mbox{for}\qquad i=0,1,2,...,m-1 \; .
\label{orbit1}
\ee

(2) The second orbit ends at the same variable position $x_0$ on the local stable manifold $M_{\rm s}(b)$ of the unstable fixed point $b$, coming from the vicinity of the attracting fixed point $a$.  This orbit is composed of the points
\be
\tilde{x}_i=\tilde{x}_i(x_0)  \qquad\mbox{for}\qquad i=m+1,...,n-1,n \qquad\mbox{with}\qquad \tilde{x}_n=x_0 \; ,
\label{orbit2}
\ee
which belong to the stable manifold $M_{\rm s}(b)$.

\begin{figure}[htbp]
\centerline{\includegraphics[width=9cm]{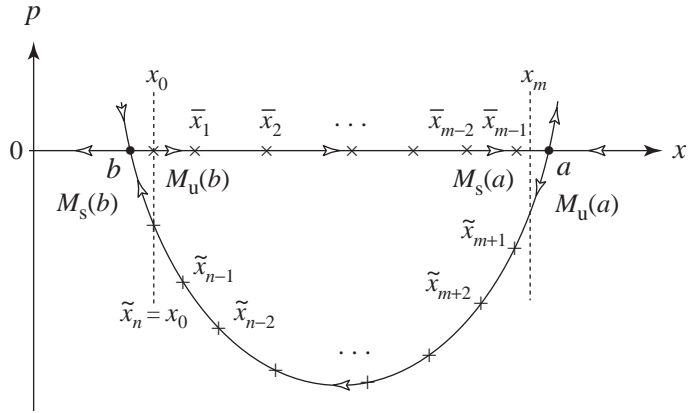}}
\caption{Schematic phase portrait of the symplectic map (\ref{sympl_map}) with the two fixed points $a$ and $b$ connected by the heteroclinic tangle formed by their stable and unstable manifolds $M_{\rm s,u}(a)$ and $M_{\rm s,u}(b)$.  The stable and unstable manifolds $M_{\rm s}(a)$ and $M_{\rm u}(b)$ belong to the axis $p=0$ where they have a common segment between the fixed points $a$ and $b$. In contrast, the stable and unstable manifolds $M_{\rm s}(b)$ and $M_{\rm u}(a)$ do not coincide but intersect to form the two heteroclinic orbits $h_{ab}^{\pm}$, which are not depicted in the figure.  The figure only shows the segment of $M_{\rm s}(b)$ from the unstable fixed point $b$ to a point of intersection with $M_{\rm u}(a)$ and this latter from this point of intersection to the attracting fixed point $a$.  For this reason, the heteroclinic tangle is not depicted here.}
\label{fig2}
\end{figure}

Therefore, the variable $x_0$ uniquely determines both orbits $\{\bar{x}_i(x_0)\}$ and $\{\tilde{x}_i(x_0)\}$.  These orbits are depicted in Fig.~\ref{fig2}.

The point $x_m$ is variable as $x_0$, but it is not attached to any of these two orbits.  Nevertheless, its contribution is expected to be the largest if it remains close to its neighbors $\bar{x}_{m-1}(x_0)$ and $\tilde{x}_{m+1}(x_0)$ since the terms with $i=m-1$ and $i=m$ in the action (\ref{action}) increase otherwise.

In the path integral (\ref{trace-int}), the variables are thus taken in the vicinity of these two orbits as
\bea
&& x_0 \; , \nonumber\\
&& x_1=\bar{x}_1(x_0) + \delta x_1 \; , \nonumber\\
&&\qquad \vdots\nonumber\\
&& x_{m-1}=\bar{x}_{m-1}(x_0) + \delta x_{m-1} \; , \nonumber\\
&& x_m \; , \label{setup}\\
&& x_{m+1}=\tilde{x}_{m+1}(x_0) + \delta x_{m+1} \; , \nonumber\\
&&\qquad \vdots\nonumber\\
&& x_{n-1}=\tilde{x}_{n-1}(x_0) + \delta x_{n-1} \; , \nonumber
\eea
with the condition
\be
x_{n}=\tilde{x}_{n}(x_0) =x_0
\ee
so that the path forms a loop.  

We notice that there exist $n$ equivalent ways to associate the integration variables with the two orbits.  We could have taken $\{x_j,x_{j+1}=f(x_j)+\delta x_{j+1},x_{j+2}=f(x_2)+\delta x_{j+2},...\}$ as well.  The choice (\ref{setup}) corresponds to $j=0$ among the $n$ possible choices $j=0,1,2,...,n-1$.  The multiple integral (\ref{trace-int}) is carried out in the whole $n$-dimensional space $(x_0,x_1,...,x_{n-1})\in{\mathbb R}^n$.  In this space, the integral evaluated in the vicinity of every one of these $n$ different choices gives the same contribution.  Therefore, the whole integral is equal to $n$ times the integral evaluated with the choice~(\ref{setup})
\bea
{\rm tr}\, \hat P^n\Big\vert_{\rm loop} &=& n \int \frac{dx_0}{\sqrt{2\pi\epsilon}} \, \frac{d\delta x_1}{\sqrt{2\pi\epsilon}} \, \cdots \frac{d\delta x_{m-1}}{\sqrt{2\pi\epsilon}} \, \frac{dx_m}{\sqrt{2\pi\epsilon}} \, \frac{d\delta x_{m+1}}{\sqrt{2\pi\epsilon}} \, \cdots \frac{d\delta x_{n-1}}{\sqrt{2\pi\epsilon}} \nonumber\\
&&\qquad\times \exp\left[-\frac{1}{\epsilon}W_n(x_0,\bar{x}_1+\delta x_1,...,\bar{x}_{m-1}+\delta x_{m-1},x_m,\tilde{x}_{m+1}+\delta x_{m+1},...,\tilde x_{n-1}+\delta x_{n-1})\right] \, .
\label{trace-int-delta}
\eea

For the purpose of carrying out the integrals over the variables $\delta x_i$ by the method of steepest descent in the weak-noise limit, the action (\ref{action}) can be expanded in these variables up to second order as
\be
W_n= W_n^{\rm orb} + \bar{\bf C}\cdot\delta\bar{\bf x} + \tilde{\bf C}\cdot\delta\tilde{\bf x}+ \frac{1}{2}\,\delta\bar{\bf x}\cdot\bar{\mbox{\helvb D}}\cdot\delta\bar{\bf x} + \frac{1}{2}\,\delta\tilde{\bf x}\cdot\tilde{\mbox{\helvb D}}\cdot\delta\tilde{\bf x} + \cdots
\ee
with
\bea
\delta\bar{\bf x} &=& \left(\delta x_1,\delta x_2,...,\delta x_{m-1}\right)\, , \\
\delta\tilde{\bf x} &=& \left(\delta x_{m+1},\delta x_{m+2},...,\delta x_{n-1}\right) \, .
\eea

$W_n^{\rm orb}$ denotes the action evaluated on the orbits (\ref{orbit1}) and (\ref{orbit2}).  Since the orbit (\ref{orbit1}) belongs to the axis $p=0$, its action is vanishing and we have
\be
W_n^{\rm orb}= \frac{1}{2} \left[ x_m-f(\bar{x}_{m-1})\right]^2 + \frac{1}{2} \left[ \tilde{x}_{m+1}-f(x_m)\right]^2 + \frac{1}{2} \left[ \tilde{x}_{m+2}-f(\tilde{x}_{m+1})\right]^2 + \cdots + \frac{1}{2} \left[ \tilde{x}_{n-1}-f(\tilde{x}_{n-2})\right]^2 + \frac{1}{2} \left[ x_0-f(\tilde{x}_{n-1})\right]^2 .
\label{action-orb}
\ee

The first-order contributions are given by
\bea
&& \bar{\bf C}\cdot\delta\bar{\bf x} = \Big\{ \underbrace{\bar{x}_{m-1}-f(\bar{x}_{m-2})}_{=\, 0} -f'(\bar{x}_{m-1})\left[x_m-f(\bar{x}_{m-1})\right]\Big\} \, \delta x_{m-1}  \, ,
\label{C-bar}\\
&& \tilde{\bf C}\cdot\delta\tilde{\bf x} = \Big\{ \tilde{x}_{m+1}-f(x_m) -f'(\tilde{x}_{m+1})\left[\tilde{x}_{m+2}-f(\tilde{x}_{m+1})\right]\Big\} \, \delta x_{m+1} \, ,
\label{C-tilde}
\eea
where the other terms are vanishing, because the orbits (\ref{orbit1}) and (\ref{orbit2}) separately satisfy the second-order recurrence~(\ref{2nd-order_map}) although the point $x_m$, to which $\bar{x}_{m-1}$ and $\tilde{x}_{m+1}$ are connected, does not belong to them.

The second-order contributions involve the $(m-1)\times(m-1)$ matrix
\be
\bar{\mbox{\helvb D}} = 
\left(
\begin{array}{llllll}
\bar{A}_1 & \bar{B}_1 & 0 & \ldots & 0 & 0 \\
\bar{B}_1 & \bar{A}_2 & \bar{B}_2 & \ldots & 0 & 0 \\
0 & \bar{B}_2 & \bar{A}_3 & \ldots & 0 & 0 \\
\vdots & \vdots & \vdots & \ddots & \vdots & \vdots \\
0 & 0 & 0 & \ldots & \bar{A}_{m-2} & \bar{B}_{m-2} \\
0 & 0 & 0 & \ldots & \bar{B}_{m-2} & \bar{A}_{m-1}
\end{array}
\right)
\label{D-bar}
\ee
where
\be
\bar{A}_i = 1 + f'(\bar{x}_i)^2 \qquad\mbox{and}\qquad \bar{B}_i = -f'(\bar{x}_i)
\ee
for $i=1,2,...,m-2$, and
\be
\bar{A}_{m-1} = 1 + f'(\bar{x}_{m-1})^2-f''(\bar{x}_{m-1}) \left[x_m-f(\bar{x}_{m-1})\right]
\ee
with the notation $f''(x)=d^2f/dx^2$.  The matrix $\bar{\mbox{\helvb D}}$ is thus evaluated along the orbit (\ref{orbit1}).

The second-order contributions also involve the $(n-m-1)\times(n-m-1)$ matrix
\be
\tilde{\mbox{\helvb D}} = 
\left(
\begin{array}{llllll}
\tilde{A}_{m+1} & \tilde{B}_{m+1}  & 0 & \ldots & 0 & 0 \\
\tilde{B}_{m+1}  & \tilde{A}_{m+2}  & \tilde{B}_{m+2} & \ldots & 0 & 0 \\
0 & \tilde{B}_{m+2} & \tilde{A}_{m+3} & \ldots & 0 & 0 \\
\vdots & \vdots & \vdots & \ddots & \vdots & \vdots \\
0 & 0 & 0 & \ldots & \tilde{A}_{n-2} & \tilde{B}_{n-2} \\
0 & 0 & 0 & \ldots & \tilde{B}_{n-2} & \tilde{A}_{n-1}
\end{array}
\right)
\label{D-tilde}
\ee
where
\be
\tilde{A}_i = 1 + f'(\tilde{x}_i)^2 - f''(\tilde{x}_i)\left[\tilde{x}_{i+1}-f(\tilde{x}_i)\right] \qquad\mbox{and}\qquad \tilde{B}_i = -f'(\tilde{x}_i)
\ee
for $i=m+1,m+2,...,n-1$. The matrix $\tilde{\mbox{\helvb D}}$ is thus evaluated along the orbit (\ref{orbit2}).

Accordingly, the trace can be written as
\be
{\rm tr}\, \hat P^n\Big\vert_{\rm loop} \simeq n \int \frac{dx_0}{\sqrt{2\pi\epsilon}} \, \frac{dx_m}{\sqrt{2\pi\epsilon}} \, \exp\left(-\frac{W_n^{\rm orb}}{\epsilon}\right) \times \bar{I} \times  \tilde{I}
\label{trace-int-delta-2-orbits}
\ee
with the integrals 
\be
\bar{I} = \int \frac{d\delta\bar{\bf x}}{(2\pi\epsilon)^{\frac{m-1}{2}}} \, \exp\left[-\frac{1}{\epsilon}\left(\bar{\bf C}\cdot\delta\bar{\bf x} + \frac{1}{2}\,\delta\bar{\bf x}\cdot\bar{\mbox{\helvb D}}\cdot\delta\bar{\bf x} \right)\right]
\label{int-delta-2-orbit1}
\ee
and
\bea
\tilde{I} = \int \frac{d\delta\tilde{\bf x}}{(2\pi\epsilon)^{\frac{n-m-1}{2}}} \, \exp\left[-\frac{1}{\epsilon}\left(\tilde{\bf C}\cdot\delta\tilde{\bf x}+ \frac{1}{2}\,\delta\tilde{\bf x}\cdot\tilde{\mbox{\helvb D}}\cdot\delta\tilde{\bf x}\right)\right]
\label{int-delta-2-orbit2}
\eea
to be evaluated over the orbits (\ref{orbit1}) and (\ref{orbit2}), respectively.

\subsection{The action is simplified}

The action is equal to zero for every orbit on the axis $p=0$.  This is the case in particular for the orbit (\ref{orbit1}), which is the piece of a heteroclinic orbit $h_{ba}$.  Therefore, the action (\ref{action-orb}) is expected to be determined by the other heteroclinic orbits $h_{ab}^{\pm}$ as $n\to\infty$.  The action (\ref{action-orb}) varies as a function of $x_0$ and takes values that are close to the actions of the heteroclinic orbits $W^h=W(h_{ab}^+) < W(h_{ab}^-)$, if $n-m-1$ is large enough and the end points of the orbit are close to the fixed points.  As a matter of fact, the actions of the heteroclinic orbits $h_{ab}^{\pm}$ are very close to each other and tend to coincide in the limit of integrability.  Indeed, the difference between these actions is equal to the area of the lobes between the stable and unstable manifolds $M_{\rm s}(b)$ and $M_{\rm u}(a)$ and this area vanishes with the chaoticity of the map~\cite{MMP84}.  In order to explicitly show the connection between Eq.~(\ref{action-orb}) and the action $W^h$ of a heteroclinic orbit, we add and subtract the actions of the pieces connecting the end points of the orbit (\ref{orbit2}) to the corresponding fixed points to get
\be
W_n^{\rm orb}=\tilde{W}(x_0) - \frac{1}{2}\sum_{i=-\infty}^{m} \left[ \tilde{x}_{i+1}-f(\tilde{x}_{i})\right]^2 - \frac{1}{2}\sum_{i=n}^{+\infty} \left[ \tilde{x}_{i+1}-f(\tilde{x}_{i})\right]^2
+ \frac{1}{2} \left[ x_m-f(\bar{x}_{m-1})\right]^2 + \frac{1}{2} \left[ \tilde{x}_{m+1}-f(x_m)\right]^2 
\label{action-orb-h}
\ee
with the action
\be
\tilde{W}(x_0)= \frac{1}{2}\sum_{i=-\infty}^{+\infty} \left[ \tilde{x}_{i+1}-f(\tilde{x}_{i})\right]^2
\label{action-pre-h}
\ee
of the orbit (\ref{orbit2}) extended towards the unstable fixed point $b$ by the points $\{\tilde{x}_i(x_0)\}_{i=n}^{+\infty}$ belonging to the local stable manifold $M_{\rm s}(b)$, and towards the attracting fixed point $a$ by the points $\{\tilde{x}_i(x_0)\}_{i=-\infty}^{m}$ belonging to the local unstable manifold $M_{\rm u}(a)$.  A condition for this construction to hold is that $x_0$ should be close enough to $b$ so that $\tilde{x}_{m+1}(x_0)<a$ for given value of $n-m-1$.  For specific values of $x_0=x_{0,k}^+$, this extended orbit coincides with the heteroclinic orbit $h_{ab}^+$ and $\tilde{W}(x_{0,k}^+)=W(h_{ab}^+)=W^h$. For intermediate values of $x_0=x_{0,k}^-$, it coincides with the other heteroclinic orbit $h_{ab}^-$ and $\tilde{W}(x_{0,k}^-)=W(h_{ab}^-)$.  Otherwise, the action (\ref{action-pre-h}) takes a value in between.  As long as the difference between $W(h_{ab}^+)$ and $W(h_{ab}^-)$ is small enough, the action (\ref{action-pre-h}) can be approximated as $\tilde{W}(x_0)\simeq W^h$ by the action of the heteroclinic orbit $h_{ab}^+$.

If $\tilde{x}_{m+1}$ is close to the attracting fixed point $a$, the first infinite sum in Eq.~(\ref{action-orb-h}) can be approximated by linearizing the symplectic map near this fixed point by Eq.~(\ref{lin_map_a}) to get
\be
\frac{1}{2}\sum_{i=-\infty}^{m} \left[ \tilde{x}_{i+1}-f(\tilde{x}_{i})\right]^2 \simeq \frac{1}{2}\left( 1-\Lambda_a^2\right) \left(\tilde{x}_{m+1}-a\right)^2 .
\label{action-lin-a}
\ee
The second infinite sum in Eq.~(\ref{action-orb-h})  can be approximated similarly if $\tilde{x}_n=x_0$ is close enough to the unstable fixed point $b$ and we get
\be
\frac{1}{2}\sum_{i=n}^{+\infty} \left[ \tilde{x}_{i+1}-f(\tilde{x}_{i})\right]^2 \simeq \frac{1}{2}\left(\Lambda_b^2-1\right) \left(\tilde{x}_n-b\right)^2 = \frac{1}{2}\left(\Lambda_b^2-1\right) \left(x_0-b\right)^2 .
\label{action-lin-b}
\ee
Accordingly, the action (\ref{action-orb-h}) becomes
\be
W_n^{\rm orb} \simeq W^h - \frac{1}{2}\left( 1-\Lambda_a^2\right) \left(\tilde{x}_{m+1}-a\right)^2
 - \frac{1}{2}\left(\Lambda_b^2-1\right) \left(x_0-b\right)^2
+ \frac{1}{2} \left[ x_m-f(\bar{x}_{m-1})\right]^2 + \frac{1}{2} \left[ \tilde{x}_{m+1}-f(x_m)\right]^2 .
\label{action-orb-h-lin}
\ee

\section{The contribution of the first heteroclinic orbit}
\label{ho1}

Here, we consider the contribution of the orbit (\ref{orbit1}) and the corresponding integral (\ref{int-delta-2-orbit1}).  This is a Gaussian integral with the quadratic part given by the matrix~(\ref{D-bar}) and the linear part by Eq.~(\ref{C-bar}).  The matrix~(\ref{D-bar}) is associated with the orbit (\ref{orbit1}) on the axis $p=0$, starting from the point $x_0$ close to the unstable fixed point $b$, and going to the point $\bar{x}_{m-1}$ close to the attracting fixed point $a$.  In the limit $m\to\infty$, this matrix would thus have the form
\be
\bar{\mbox{\helvb D}} \simeq
\left(
\begin{array}{ccccc}
\ddots & \vdots & \vdots & \vdots & \vdots \\
\ldots & 1+\Lambda_a^2 & -\Lambda_a & 0 & 0 \\
\ldots & -\Lambda_a & 1+\Lambda_a^2 & -\Lambda_a & 0 \\
\ldots & 0 & -\Lambda_a & 1+\Lambda_a^2 & -\Lambda_a \\
\ldots & 0 & 0 & -\Lambda_a & 1+\Lambda_a^2 + X
\end{array}
\right)
\label{D-bar-infty}
\ee
where $X=O(x_m-a)$.  If this $(m-1)\times(m-1)$ matrix was uniformly filled as shown, its determinant would be equal to
\be
\det\bar{\mbox{\helvb D}} \simeq \frac{1-\Lambda_a^{2m}}{1-\Lambda_a^2} + X \, \frac{1-\Lambda_a^{2m-2}}{1-\Lambda_a^2}
\ee
while the last diagonal matrix element of its inverse would be given by
\be
\left(\bar{\mbox{\helvb D}}^{-1}\right)_{m-1,m-1} \simeq \frac{1-\Lambda_a^{2m-2}}{1-\Lambda_a^{2m}+X(1-\Lambda_a^{2m-2})} \; .
\ee
Since $\vert\Lambda_a\vert <1$, the limit $m\to\infty$ of these quantities would behave as
\bea
\det\bar{\mbox{\helvb D}} &\simeq& \frac{1}{1-\Lambda_a^2} + O(x_m-a) \; ,\label{detD-bar-infty} \\
\left(\bar{\mbox{\helvb D}}^{-1}\right)_{m-1,m-1} &\simeq& 1+O(x_m-a) \; .\label{invD-bar-infty}
\eea
Given that the point $x_m$ is linked to the points $\bar{x}_{m-1}$ and $\tilde{x}_{m+1}$, which are arbitrarily close to the attracting fixed point $a$, the correction $O(x_m-a)$ may be supposed negligible in the limit $m\to\infty$.  It turns out that, even if the matrix is asymptotically of the form shown in Eq.~(\ref{D-bar-infty}), the determinant and the last diagonal element of its inverse are still given by Eqs.~(\ref{detD-bar-infty})-(\ref{invD-bar-infty}).

We notice that the matrix (\ref{D-bar}) evaluated along the orbit (\ref{orbit1}) has an eigenvalue that is vanishing in the limit $m\to\infty$, which corresponds to the zero mode of the infinite matrix associated with the heteroclinic orbit $h_{ba}$.  Indeed, as aforementioned, such a heteroclinic orbit belongs to a continuous family of heteroclinic orbits obtained by varying the starting point $x_0\in[b,a]$.  Differentiating Eq.~(\ref{2nd-order_map}) with respect to the initial condition $x_0$ shows that the infinite matrix admits a zero eigenvalue associated with this zero mode of translation of the heteroclinic orbit with respect to $x_0$.  In the finite matrix (\ref{D-bar}), this zero eigenvalue corresponds to a vanishing eigenvalue, which is the smallest of this matrix. In spite of the existence of this vanishing eigenvalue, the determinant does not vanish but approaches the finite value given by Eq.~(\ref{detD-bar-infty}) because  the other non-vanishing eigenvalues compensate the presence of the vanishing one.

Accordingly, the integral (\ref{int-delta-2-orbit1}) is evaluated as
\be
\bar{I} = \frac{1}{\sqrt{\det\bar{\mbox{\helvb D}}}} \, \exp\left(\frac{1}{2\epsilon}\bar{\bf C}\cdot\bar{\mbox{\helvb D}}^{-1}\cdot\bar{\bf C} \right) \, .
\ee
By using Eqs.~(\ref{C-bar}) and (\ref{invD-bar-infty}), we get
\be
\bar{\bf C}\cdot\bar{\mbox{\helvb D}}^{-1}\cdot\bar{\bf C}=\left(\bar{\mbox{\helvb D}}^{-1}\right)_{m-1,m-1} \Big\{ -f'(\bar{x}_{m-1})\left[x_m-f(\bar{x}_{m-1})\right]\Big\}^2 \simeq \Lambda_a^2\, (x_m-a)^2 + O\left[(x_m-a)^3\right] .
\label{CDC-bar}
\ee
With Eq.~(\ref{detD-bar-infty}), the integral is thus evaluated as
\be
\bar{I} \simeq \sqrt{1-\Lambda_a^2} \, \exp\left(-\frac{\Delta\bar{W}}{\epsilon}\right)
\qquad\mbox{with}\qquad
\Delta\bar{W}=-\frac{1}{2}\, \Lambda_a^2\, (x_m-a)^2
\label{I-bar+DW-bar}
\ee
up to corrections of higher orders in $O(x_m-a)$.

\section{The contribution of the second heteroclinic orbit}
\label{ho2}

Now, we turn to the evaluation of the integral (\ref{int-delta-2-orbit2}) around the orbit (\ref{orbit2}).  As for the previous matrix (\ref{D-bar}), the matrix (\ref{D-tilde}) depends on $x_0$ and has a vanishing eigenvalue.  However, contrary to the situation with the previous matrix, the vanishing eigenvalue of the matrix (\ref{D-tilde}) oscillates between small positive and negative values as $x_0$ is varied and the determinant of the matrix (\ref{D-tilde}) changes sign accordingly, because the other eigenvalues are positive.  Consequently, the integration should be carried out by treating separately the zero mode associated with the vanishing eigenvalue.

The matrix (\ref{D-tilde}) is real symmetric and can thus be diagonalized as
\be
\tilde{\mbox{\helvb D}} = \mbox{\helvb O}^{\rm T}\cdot\pmb{\lambda}\cdot\mbox{\helvb O}
\ee
with an orthogonal transformation $\mbox{\helvb O}$ such that  $\mbox{\helvb O}^{-1}=\mbox{\helvb O}^{\rm T}$ and $\vert\det\mbox{\helvb O}\vert=1$.  The diagonal matrix $\pmb{\lambda}=(\lambda_{\nu}\,\delta_{\nu\nu'})$ contains the eigenvalues of the matrix (\ref{D-tilde}) on its diagonal.  The associated eigenvectors
\be
\tilde{\mbox{\helvb D}}\cdot{\bf v}_{\nu} = \lambda_{\nu} \, {\bf v}_{\nu}
\ee
form an orthonormal basis ${\bf v}_{\nu}\cdot{\bf v}_{\nu'}=\delta_{\nu\nu'}$ since their components are the elements of the orthogonal matrix
\be
({\bf v}_{\nu})_i= \left(\mbox{\helvb O}\right)_{\nu i} = \left(\mbox{\helvb O}^{\rm T}\right)_{i \nu}  .
\ee
We introduce the new integration variables
\be
\delta{\bf y}=\mbox{\helvb O}\cdot\delta\tilde{\bf x} \; .
\ee
As long as the small eigenvalue $\lambda_0$ is not equal to zero, the integral (\ref{int-delta-2-orbit2}) becomes
\be
\tilde{I} = \int \frac{d\delta y_0}{\sqrt{2\pi\epsilon}}\, \exp\left[-\frac{\lambda_0}{2\epsilon}\left(\delta y_0 + \frac{\tilde c_0}{\lambda_0}\right)^2\right] \, \sqrt{\frac{\lambda_0}{\det\tilde{\mbox{\helvb D}}}} \, \exp\left(\frac{1}{2\epsilon}\tilde{\bf C}\cdot\tilde{\mbox{\helvb D}}^{-1}\cdot\tilde{\bf C} \right)
\label{I-tilde-y0}
\ee
where $\tilde{c}_0=\tilde{\bf C}\cdot{\bf v}_0$ and $\det\tilde{\mbox{\helvb D}}=\prod_{\nu}\lambda_{\nu}$.

The remaining integral is carried out in the direction $\delta y_0={\bf v}_0\cdot\delta\tilde{\bf x}$ of the eigenvector ${\bf v}_0$, which is associated with translations along the stable manifold $M_{\rm s}(b)$ of the unstable fixed point $b$.  Locally near the unstable fixed point, the stable manifold has the parametric equations
\be
M_{\rm s}^{\rm loc}(b)\left\{
\begin{array}{l}
\delta x(\tau) = \frac{\delta x_0}{\Lambda_b^{\tau}} \\
\delta p(\tau) =(1-\Lambda_b^2)\, \delta x(\tau) = (1-\Lambda_b^2)\, \frac{\delta x_0}{\Lambda_b^{\tau}}
\end{array}
\right.
\ee
with $\delta x(0)=\delta x_0=x_0-b$ and $\delta x(\tau)=x(\tau)-b$.  The parameter $\tau$ varies continuously.  The points of the orbit (\ref{orbit2}) that are close to the unstable fixed point $b$ are given by $\tilde x_i=\tilde x_i(x_0) = b+\delta x(\tau)$ with $\tau=i-n$.  Integrating over $\delta y_0$ is equivalent to integrating over the parameter $\tau\in[0,1[$, which corresponds to translations along the stable manifold $M_{\rm s}(b)$.  Thanks to this translational degree of freedom, the constant $\tilde c_0/\lambda_0$ can be eliminated by the change $\delta y'_0 = \delta y_0+\tilde c_0/\lambda_0$.  Moreover, since $\lambda_0$ is vanishingly small, the first Gaussian function in Eq.~(\ref{I-tilde-y0}) is well approximated by the unit value.

The change of integration variable from $\delta y_0$ to $\tau\in[0,1[$ involves the following Jacobian factor
\be
d\delta y_0 = \left\vert {\bf v}_0\cdot\frac{d\tilde{\bf x}}{d\tau}\right\vert \, d\tau \; .
\label{Jacob}
\ee
Therefore, the integral (\ref{I-tilde-y0}) becomes
\be
\tilde{I} \simeq \int_0^1 \frac{d\tau}{\sqrt{2\pi\epsilon}}\, \left\vert {\bf v}_0\cdot\frac{d\tilde{\bf x}}{d\tau}\right\vert \, \sqrt{\frac{\lambda_0}{\det\tilde{\mbox{\helvb D}}}} \, \exp\left(\frac{1}{2\epsilon}\tilde{\bf C}\cdot\tilde{\mbox{\helvb D}}^{-1}\cdot\tilde{\bf C} \right) .
\label{I-tilde-tau}
\ee

By using Eq.~(\ref{C-tilde}), we have that
\be
\tilde{\bf C}\cdot\tilde{\mbox{\helvb D}}^{-1}\cdot\tilde{\bf C}=\left(\tilde{\mbox{\helvb D}}^{-1}\right)_{m+1,m+1} \Big\{ \tilde{x}_{m+1}-f(x_m) -f'(\tilde{x}_{m+1})\left[\tilde{x}_{m+2}-f(\tilde{x}_{m+1})\right]\Big\}^2 .
\ee
Interestingly, the inverse of the matrix (\ref{D-tilde}) is also evaluated close to the attracting fixed point where it behaves as
\be
\tilde{\mbox{\helvb D}} \simeq
\left(
\begin{array}{ccccc}
1+\Lambda_a^2 & -\Lambda_a & 0 & 0 & \ldots \\
-\Lambda_a & 1+\Lambda_a^2 & -\Lambda_a & 0 & \ldots\\
0 & -\Lambda_a & 1+\Lambda_a^2 & -\Lambda_a & \ldots\\
0 & 0 & -\Lambda_a & 1+\Lambda_a^2 & \ldots\\
\vdots & \vdots & \vdots & \vdots &\ddots \\
\end{array}
\right)
\label{D-tilde-infty-a}
\ee
so that
\be
\left(\bar{\mbox{\helvb D}}^{-1}\right)_{m+1,m+1} \simeq 1+O\left(\Lambda_a^{2(n-m)}\right) .
\ee
Accordingly, the integral (\ref{I-tilde-y0}) can be written as
\be
\tilde{I} \simeq \int_0^1 \frac{d\tau}{\sqrt{2\pi\epsilon}}\, \left\vert {\bf v}_0\cdot\frac{d\tilde{\bf x}}{d\tau}\right\vert \, \sqrt{\frac{\lambda_0}{\det\tilde{\mbox{\helvb D}}}} \, \exp\left(-\frac{\Delta\tilde{W}}{\epsilon}\right)
\label{I-tilde-tau-bis}
\ee
with
\be
\Delta\tilde{W} = - \frac{1}{2}\, \Big\{ \tilde{x}_{m+1}-f(x_m) -f'(\tilde{x}_{m+1})\left[\tilde{x}_{m+2}-f(\tilde{x}_{m+1})\right]\Big\}^2 .
\label{DW-tilde}
\ee

Close to the unstable fixed point $b$, the matrix (\ref{D-tilde}) behaves as
\be
\tilde{\mbox{\helvb D}} \simeq
\left(
\begin{array}{ccccc}
\ddots & \vdots & \vdots & \vdots & \vdots \\
\ldots & 1+\Lambda_b^2 & -\Lambda_b & 0 & 0 \\
\ldots & -\Lambda_b & 1+\Lambda_b^2 & -\Lambda_b & 0 \\
\ldots & 0 & -\Lambda_b & 1+\Lambda_b^2 & -\Lambda_b \\
\ldots & 0 & 0 & -\Lambda_b & 1+\Lambda_b^2
\end{array}
\right) .
\label{D-tilde-infty-b}
\ee
Consequently, the eigenvector ${\bf v}_0$ associated with the vanishing eigenvalue $\lambda_0$ is the solution of a second-order recurrence with constant coefficients so that
\be
v_{0,i} = \frac{A}{\Lambda_b^{i-n}}+ B \, \Lambda_b^{i-n} \qquad\mbox{for}\qquad i=...,n-2,n-1
\ee
where $A$ and $B$ are two constants to be fixed by the boundary condition $v_{0,n}=0$ and the value of $v_{0,n-1}$ for instance.  Therefore, we find
\be
v_{0,i} = v_{0,n-1} \frac{\Lambda_b}{\Lambda_b^2-1} \left(\frac{1}{\Lambda_b^{i-n}}-\Lambda_b^{i-n}\right) \qquad\mbox{for}\qquad i\leq n \; .
\label{v0i}
\ee

This behavior should be compared with the zero mode $d\tilde{\bf x}/d\tau$ introduced in Eq.~(\ref{Jacob}) as the derivative of the orbit~(\ref{orbit2}) with respect to the translation parameter $\tau$.
We notice that the derivatives with respect to $\tau$ and $x_0$ are related to each other.
Since $x_0\simeq b+K \exp(-\tau \ln\Lambda_b)$, changing $\tau$ by $d\tau$ is indeed equivalent to varying $x_0$ by 
\be
dx_0 = -(x_0-b) \, \ln\Lambda_b \, d\tau \; .
\label{x0-tau}
\ee
Away from the ends of the orbit, we expect a proportionality between the eigenvector ${\bf v}_0$ and the zero mode $d\tilde{\bf x}/d\tau$.  The reason is that the orbit (\ref{orbit2}) obeys the second-order recurrence (\ref{2nd-order_map}).  If we differentiate Eq.~(\ref{2nd-order_map}) with respect to $x_0$ or $\tau$, we get 
\be
\tilde{\mbox{\helvb D}}_{\infty}\cdot\frac{d\tilde{\bf x}}{d\tau}=0
\ee
where $\tilde{\mbox{\helvb D}}_{\infty}$ is the infinite matrix obtained by extending the finite matrix (\ref{D-tilde}) towards the future and the past of the orbit (\ref{orbit2}).  $d\tilde{\bf x}/d\tau$ is called the zero mode because it is the eigenvector associated with the zero eigenvalue of this infinite matrix.  Close to the unstable fixed point $b$, the zero mode behaves as
\be
\frac{d\tilde x_i}{d\tau} = \frac{1}{\Lambda_b^{i-n}} \, \frac{d\tilde x_n}{d\tau} \qquad\mbox{for}\qquad i\lesssim n
\label{zm-exp}
\ee
and Eq.~(\ref{v0i}) shows that the eigenvector ${\bf v}_0$ has a similar behavior 
\be
v_{0,i} \simeq  v_{0,n-1} \frac{\Lambda_b}{\Lambda_b^2-1} \, \frac{1}{\Lambda_b^{i-n}}
 \qquad\mbox{for}\qquad i\lesssim n \; .
\label{v0-exp}
\ee
Taking the ratio between Eqs.~(\ref{zm-exp}) and (\ref{v0-exp}), we find that
\be
\frac{d\tilde x_i}{d\tau} \simeq \frac{d\tilde x_n}{d\tau}\, \frac{\Lambda_b^2-1}{\Lambda_b} \, \frac{v_{0,i}}{v_{0,n-1}} \; .
\ee
Since $\tilde x_n=x_0$, we get from Eq.~(\ref{x0-tau}) that
\be
\frac{d\tilde x_n}{d\tau} = -(x_0-b) \, \ln\Lambda_b 
\ee
whereupon
\be
\frac{d\tilde{\bf x}}{d\tau} \simeq -(x_0-b) \, \ln\Lambda_b \, \frac{\Lambda_b^2-1}{\Lambda_b} \, \frac{{\bf v}_0}{v_{0,n-1}}
\ee
and
\be
\left\vert{\bf v}_0\cdot\frac{d\tilde{\bf x}}{d\tau}\right\vert \simeq (x_0-b) \, \frac{\Lambda_b^2-1}{\Lambda_b} \, \frac{\ln\Lambda_b}{\vert v_{0,n-1}\vert}
\label{expression}
\ee
since the eigenvector ${\bf v}_0$ is normalized to unity, ${\bf v}_0^2=1$.  The right-hand side of Eq.~(\ref{expression}) is positive because $x_0>b$ and $\Lambda_b>1$.

Finally, the integral (\ref{I-tilde-tau-bis}) is evaluated as
\be
\tilde{I} \simeq \frac{x_0-b}{\sqrt{2\pi\epsilon}}\, (\Lambda_b^2-1) \, \frac{\ln\Lambda_b}{\Lambda_b} \, \sqrt{\frac{\lambda_0}{v_{0,n-1}^2 \det\tilde{\mbox{\helvb D}}}} \, \exp\left(-\frac{\Delta\tilde{W}}{\epsilon}\right)
\label{I-tilde-tau-ter}
\ee
where $\Delta\tilde{W}$ is given by Eq.~(\ref{DW-tilde}).  

As confirmed by the numerical analysis of the examples presented in Section~\ref{Examples}, the expression in the square root is well defined if $n-m-1$ is large enough and $x_0-b$ small enough.  On the one hand, the ratio $\lambda_0/\det\tilde{\mbox{\helvb D}}$ is well defined when the eigenvalue $\lambda_0$ vanishes, because $\det\tilde{\mbox{\helvb D}}$ is proportional to $\lambda_0$. On the other hand, the end $v_{0,n-1}$ of the eigenvector ${\bf v}_0$ is going to zero as $v_{0,n-1}=O(x_0-b)\to 0$, but the determinant diverges as $\det\tilde{\mbox{\helvb D}}=O\left[(x_0-b)^{-2}\right]$ in the same limit.  Therefore, the product $v_{0,n-1}^2\det\tilde{\mbox{\helvb D}}$ is well defined in the limit $x_0\to b$.  Under these conditions, the expression in the square root is thus a well-defined quantity.

\section{From the trace to the escape rate}
\label{Rate}

\subsection{The contributions of both orbits in the loop}

Collecting the results (\ref{I-bar+DW-bar}), (\ref{DW-tilde}), and~(\ref{I-tilde-tau-ter}), the contribution of the loop to the trace (\ref{trace-int-delta-2-orbits}) becomes
\be
{\rm tr}\, \hat P^n\Big\vert_{\rm loop} \simeq n \int \frac{dx_0}{\sqrt{2\pi\epsilon}} \, \frac{dx_m}{\sqrt{2\pi\epsilon}} \, \sqrt{1-\Lambda_a^2} \, \frac{x_0-b}{\sqrt{2\pi\epsilon}}\, (\Lambda_b^2-1) \, \frac{\ln\Lambda_b}{\Lambda_b} \, \sqrt{\frac{\lambda_0}{v_{0,n-1}^2 \det\tilde{\mbox{\helvb D}}}} \, \exp\left(-\frac{W_n}{\epsilon}\right)
\label{trace-int-delta-loop}
\ee
with the action including the additional terms from the boundaries of the heteroclinic orbits
\bea
W_n &=&W_n^{\rm orb}+\Delta\bar{W}+\Delta\tilde{W} \nonumber\\
&\simeq& W^h - \frac{1}{2}\left( 1-\Lambda_a^2\right) \left(\tilde{x}_{m+1}-a\right)^2
 - \frac{1}{2}\left(\Lambda_b^2-1\right) \left(x_0-b\right)^2
+ \frac{1}{2} \left[ x_m-f(\bar{x}_{m-1})\right]^2 + \frac{1}{2} \left[ \tilde{x}_{m+1}-f(x_m)\right]^2 \nonumber\\
&&\qquad
-\frac{1}{2}\, \Lambda_a^2\, (x_m-a)^2
 - \frac{1}{2}\, \Big\{ \tilde{x}_{m+1}-f(x_m) -f'(\tilde{x}_{m+1})\left[\tilde{x}_{m+2}-f(\tilde{x}_{m+1})\right]\Big\}^2 .
\eea

\subsection{The action is further simplified}

Since the end point $\bar{x}_{m-1}$ of the orbit (\ref{orbit1}) is close to the attracting fixed point $a$, we can use the approximation
\be
\frac{1}{2} \left[ x_m-f(\bar{x}_{m-1})\right]^2 \simeq \frac{1}{2} \left( x_m-a\right)^2
\ee
and the action can be written as
\bea
W_n &\simeq& W^h - \frac{1}{2}\left(\Lambda_b^2-1\right) \left(x_0-b\right)^2
+ \frac{1}{2} \left(1-\Lambda_a^2\right) (x_m-a)^2 \nonumber\\
&&\qquad
- \frac{1}{2}\left( 1-\Lambda_a^2\right) \left(\tilde{x}_{m+1}-a\right)^2
+ \frac{1}{2} \left[ \tilde{x}_{m+1}-f(x_m)\right]^2 \nonumber\\
&&\qquad - \frac{1}{2}\, \Big\{ \tilde{x}_{m+1}-f(x_m) -f'(\tilde{x}_{m+1})\left[\tilde{x}_{m+2}-f(\tilde{x}_{m+1})\right]\Big\}^2 .
\eea

Using the second-order recurrence (\ref{2nd-order_map}) with $i=m+1$ for the orbit (\ref{orbit2})
\be
\tilde{x}_{m+1}-f(\tilde{x}_{m})-f'(\tilde{x}_{m+1})\left[ \tilde{x}_{m+2}-f(\tilde{x}_{m+1})\right]=0 \; ,
\ee
the last term of the action is simplified and we get
\bea
W_n &\simeq& W^h - \frac{1}{2}\left(\Lambda_b^2-1\right) \left(x_0-b\right)^2
+ \frac{1}{2} \left(1-\Lambda_a^2\right) (x_m-a)^2 \nonumber\\
&&\qquad
- \frac{1}{2}\left( 1-\Lambda_a^2\right) \left(\tilde{x}_{m+1}-a\right)^2
+ \frac{1}{2} \left[ \tilde{x}_{m+1}-f(x_m)\right]^2  - \frac{1}{2}\, \left[ f(\tilde{x}_m)-f(x_m)\right]^2 .
\eea
Since the beginning of the orbit (\ref{orbit2}) is also close to the attracting fixed point $a$, we have that $\tilde{x}_m\simeq a$, $f(\tilde{x}_m)\simeq a$, and $\tilde{x}_{m+1}\simeq a$, whereupon the last three terms vanish and the action is finally given by
\be
W_n \simeq W^h - \frac{1}{2}\left(\Lambda_b^2-1\right) \left(x_0-b\right)^2
+ \frac{1}{2} \left(1-\Lambda_a^2\right) (x_m-a)^2 \, .
\label{W_n}
\ee

\subsection{The integrals over $x_0$ and $x_m$}

Replacing the action (\ref{W_n}) in the expression (\ref{trace-int-delta-loop}) for the trace, we obtain
\bea
{\rm tr}\, \hat P^n\Big\vert_{\rm loop} &\simeq& n \, \frac{\ln\Lambda_b}{\Lambda_b} \, \sqrt{\frac{\lambda_0}{v_{0,n-1}^2 \det\tilde{\mbox{\helvb D}}}} \, \exp\left(-\frac{W^h}{\epsilon}\right) \nonumber\\
&&\qquad\times \int\frac{dx_m}{\sqrt{2\pi\epsilon}} \, \sqrt{1-\Lambda_a^2} \, \exp\left[-\frac{1}{2\epsilon}\left(1-\Lambda_a^2\right) (x_m-a)^2 \right] \nonumber\\
&&\qquad\times \int \frac{dx_0}{2\pi\epsilon} \, (\Lambda_b^2-1)\, \left(x_0-b\right) \, \exp\left[+\frac{1}{2\epsilon}\left(\Lambda_b^2-1\right) \left(x_0-b\right)^2\right]
\label{trace-int-delta-loop-x0-xm}
\eea
where the integral over $x_m$ is carried out near the attracting fixed point $a$ and the integral over $x_0$ extends from the unstable fixed point $b$ up to the attracting fixed point $a$.

The integral over $x_m$ is a simple Gaussian equal to unity
\be
\int\frac{dx_m}{\sqrt{2\pi\epsilon}} \, \sqrt{1-\Lambda_a^2} \, \exp\left[-\frac{1}{2\epsilon}\left(1-\Lambda_a^2\right) (x_m-a)^2 \right] = 1
\ee
because $a$ is an attracting fixed point so that $\vert\Lambda_a\vert < 1$.

In contrast, the integral over $x_0$ is diverging.  This integral is known to appear in the path-integral approach to diffusion in a bistable potential \cite{CCR81} or nucleation-rate theory \cite{S81}.  The regularization of this divergence can here be performed with a rotation by $\pi/2$ in the complex plane of the variable $x_0$ around the point $b$ and, thus, by considering the new integration variable:
\be
z=-\frac{1}{2\epsilon} \, \left(\Lambda_b^2-1\right) \, \left(x_0-b\right)^2 \, .
\ee
Consequently, the integral over $x_0$ is evaluated as
\be
\int_b^{b+i\infty} \frac{dx_0}{2\pi\epsilon} \, (\Lambda_b^2-1)\, \left(x_0-b\right) \, \exp\left[+\frac{1}{2\epsilon}\left(\Lambda_b^2-1\right) \left(x_0-b\right)^2\right] = -\frac{1}{2\pi} \int_0^{\infty} dz \, {\rm e}^{-z} = -\frac{1}{2\pi} \, .
\ee
We notice that the minus sign is expected because the loop should have a negative contribution to the trace according to Eq.~(\ref{trace-exp-Taylor}).

After these integrations, the trace (\ref{trace-int-delta-loop-x0-xm}) is finally given by
\be
{\rm tr}\, \hat P^n\Big\vert_{\rm loop} \simeq - n \, \frac{\ln\Lambda_b}{2\pi\Lambda_b} \, \sqrt{\frac{\lambda_0}{v_{0,n-1}^2 \det\tilde{\mbox{\helvb D}}}} \, \exp\left(-\frac{W^h}{\epsilon}\right) .
\label{trace-int-delta-loop-fin}
\ee

\subsection{The escape rate}

Comparing Eq.~(\ref{trace-int-delta-loop-fin}) with Eq.~(\ref{trace-exp-Taylor}) allows us to identify the escape rate as
\be
\gamma \simeq \frac{\ln\Lambda_b}{2\pi\Lambda_b} \, \sqrt{\frac{\lambda_0}{v_{0,n-1}^2 \det\tilde{\mbox{\helvb D}}}} \, \exp\left(-\frac{W^h}{\epsilon}\right)
\label{escape-rate-fin}
\ee
where $\Lambda_b>1$ is the stretching factor of the linearized map near the unstable fixed point $b$, $W^h$ is the action of the heteroclinic orbit $h_{ab}^+$, and the $(n-m-1)\times(n-m-1)$ matrix $\tilde{\mbox{\helvb D}}$ is evaluated in the limit $\vert n-m\vert\gg 1$ on this heteroclinic orbit together with its determinant $\det\tilde{\mbox{\helvb D}}$, its vanishing eigenvalue $\lambda_0$, and its associated eigenvector ${\bf v}_0=(v_{0,i})_{i=m+1}^{n-1}$ for its element $v_{0,n-1}$ closest to the unstable fixed point $b$.  This expression is our final result for the escape rate of the noisy map.

\section{Examples}
\label{Examples}

In the present section, we apply the formula (\ref{escape-rate-fin}) to two different noisy maps with an attracting fixed point and an unstable one.  This formula is compared with the escape rate obtained by Monte Carlo simulations, in which random trajectories of the noisy map (\ref{noisy_map}) are generated starting from the attracting fixed point and escaping when the point $x_n$ is far enough from the interval $[b,a]$.  The criterion for escape is taken as $\vert x_n\vert>d=100$ such that $d=100\gg \vert a\vert, \vert b\vert$.  The histogram of escape times is obtained for such a statistical ensemble of trajectories.  After transients, the histogram shows an exponential decay at a well-defined escape rate $\gamma$.  The computation of the escape rate is repeated for six different values of the noise amplitude $\epsilon$ in order to determine the constant $A$ and $W$ in Eq.~(\ref{Arrhenius}).  The result is compared with the theoretical value given by the formula (\ref{escape-rate-fin}).  The values are also compared with another theoretical estimation obtained in Appendix~\ref{AppB} in the limit where both fixed points coincide $a\to b$ and the noisy map is well approximated by a continuous-time Langevin stochastic process.

\subsection{The noisy logistic map}
\label{logistic}

\begin{figure}[h]
\begin{center}
\includegraphics[scale=0.46]{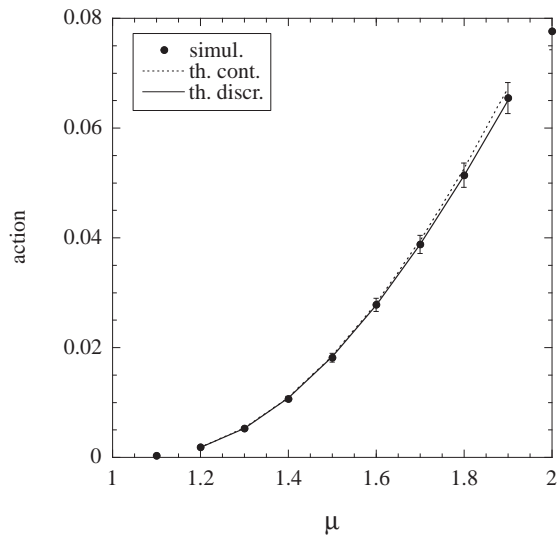}
\caption{Action $W$ of the escape rate $\gamma=A\exp(-W/\epsilon)$ versus the parameter $\mu$ for the logistic map (\ref{logistic_map}).  The dots are showing the results of Monte Carlo simulations, the solid line the theoretical result of the trace formula (\ref{escape-rate-fin}), and the dashed line the theoretical estimation (\ref{esc_rate-logistic_map}) of the continuous limit around $\mu=1$.  The estimated relative error is 4.3\%.}
\label{fig3}
\end{center}
\end{figure}

\begin{figure}[h]
\begin{center}
\includegraphics[scale=0.46]{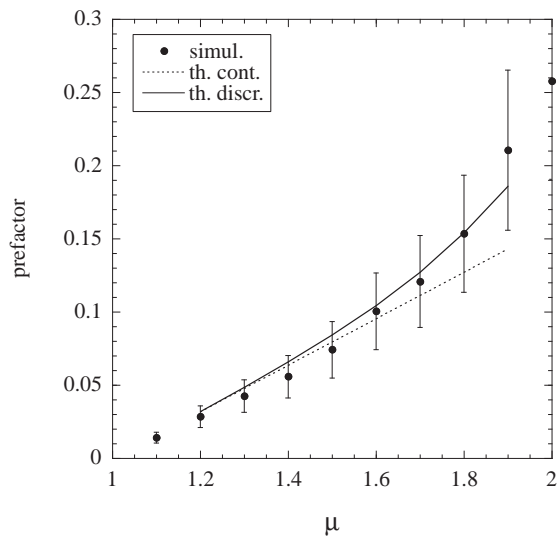}
\caption{Prefactor $A$ of the escape rate $\gamma=A\exp(-W/\epsilon)$ versus the parameter $\mu$ for the logistic map.  The dots are showing the results of Monte Carlo simulations, the solid line the theoretical result of the trace formula (\ref{escape-rate-fin}), and the dashed line the theoretical estimation (\ref{esc_rate-logistic_map}) of the continuous limit around $\mu=1$.  The estimated relative error is 26\%.}
\label{fig4}
\end{center}
\end{figure}

Let us consider the noisy logistic map (\ref{noisy_map}) with
\be 
f(x) = \mu \, x \, (1-x)
\label{logistic_map}
\ee
where $\mu$ is the control parameter.  The attracting and unstable fixed points and their associated stretching factor are given by $a=1-1/\mu$ with $\Lambda_a=2-\mu$ and $b=0$ with $\Lambda_b=\mu$ for this map.  The theoretical results are compared to Monte Carlo simulations in Figs.~\ref{fig3}-\ref{fig4} where we see their agreement.   

\subsection{The noisy exponential map}
\label{expon}

\begin{figure}[h]
\begin{center}
\includegraphics[scale=0.46]{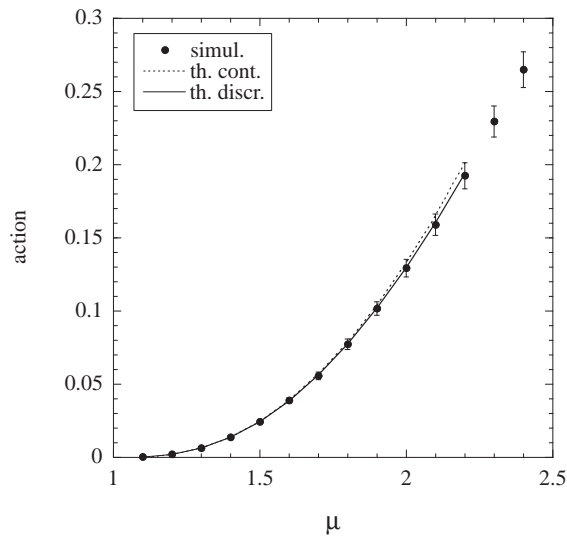}
\caption{Action $W$ of the escape rate $\gamma=A\exp(-W/\epsilon)$ versus the parameter $\mu$ for the exponential map.  The dots are showing the results of Monte Carlo simulations, the solid line the theoretical result of the trace formula (\ref{escape-rate-fin}), and the dashed line the theoretical estimation (\ref{esc_rate-exp_map}) of the continuous limit around $\mu=1$.  The estimated relative error is 4.6\%.}
\label{fig5}
\end{center}
\end{figure}

\begin{figure}[h]
\begin{center}
\includegraphics[scale=0.46]{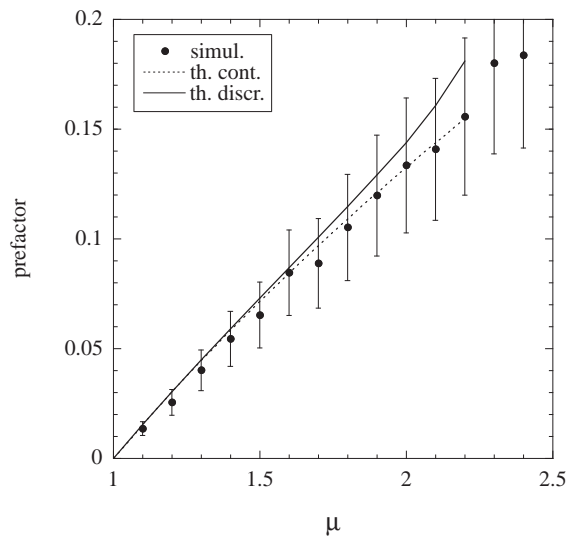}
\caption{Prefactor $A$ of the escape rate $\gamma=A\exp(-W/\epsilon)$ versus the parameter $\mu$ for the exponential map. The dots are showing the results of Monte Carlo simulations, the solid line the theoretical result of the trace formula (\ref{escape-rate-fin}), and the dashed line the theoretical estimation (\ref{esc_rate-exp_map}) of the continuous limit around $\mu=1$.  The estimated relative error is 23\%.}
\label{fig6}
\end{center}
\end{figure}

As a second example, we consider the noisy exponential map (\ref{noisy_map}) with
\be 
f(x) = \mu \, x \, \exp(-x) \, .
\label{exp_map}
\ee
For this map, we have that $a=\ln\mu$ with $\Lambda_a=1-\ln\mu$ and $b=0$ with $\Lambda_b=\mu$.  The theoretical results are compared to Monte Carlo simulations in Figs.~\ref{fig5}-\ref{fig6}, showing their agreement.  

\section{Conclusions}
\label{Conclusions}

In this paper, we have obtained a formula for the rate of escape from an attracting fixed point in one-dimensional noisy maps.  

The calculation uses path-integral methods starting from the trace of iterates of the Frobenius-Perron operator ruling the time evolution of the probability density.  In the weak-noise limit, the path integral giving this trace is dominated by the contributions of the closed orbits of a two-dimensional symplectic map, which is associated with the one-dimensional noisy map.  For this symplectic map, the attracting and unstable fixed points are connected by two heteroclinic orbits.  They form a loop, on which periodic orbits of arbitrarily large periods accumulate.  In the weak-noise limit, the loop of heteroclinic orbits is shown to determine the escape rate.  This latter is exponentially small in the noise amplitude $\epsilon$ and has the expression $\gamma\simeq A\exp(-W/\epsilon)$ characteristic of activated processes.  The constant $W$ is equal to the action of the main heteroclinic orbit from the attracting to the unstable fixed point and the prefactor $A$ is obtained in terms of the linearized symplectic map around this heteroclinic orbit and the  fixed points it connects.

The values of the formula are compared with the escape rate computed with Monte Carlo simulations in the noisy logistic and exponential maps.  The results are also compared with the escape rate calculated in the limit where the two fixed points coincide and the noisy map can be approximated by a stochastic differential equation of Langevin type.  The different results are in agreement.

The method used to obtain the formula for the escape rate is direct in the sense that the path integral is carried out step by step from the trace of the stochastic evolution operator up to the escape rate.  The method could be extended to other situations, such as escape from a periodic attractor.  The theory also shows that the escape rate of activated processes can be deduced from the trace formula along lines that are similar as in the cases where escape already manifests itself without noise \cite{KT84,CE91,G98,D99,CSPVD99,PVV01}, but with differences coming from the accumulation of periodic orbits on the heteroclinic loop.

\begin{acknowledgments}
J. Demaeyer thanks the Physics Department of Universit\'e Libre de Bruxelles for financial support.  This research is also supported by the Belgian Federal Government under the Interuniversity Attraction Pole project P7/18 ``DYGEST".
\end{acknowledgments}

\appendix

\section{Noisy linear maps}
\label{AppA}

Here, we consider simple processes ruled by the random recurrence
\be
x_{n+1}= \Lambda \, x_n + \xi_n
\ee
where $\Lambda$ is the slope of the linear map and $\xi_n$ are independent Gaussian random variables of zero mean and variance~$\epsilon$. The kernel (\ref{kernel}) of the Frobenius-Perron operator is here given by
\be
K(x_0,x) = \frac{1}{\sqrt{2\pi\epsilon}} \; \exp\left[-\frac{1}{2\epsilon}(x-\Lambda x_0)^2\right]
\ee
and its iterate by
\be
K_n(x_0,x) = \sqrt{\frac{1-\Lambda^2}{2\pi\epsilon (1-\Lambda^{2n})}} \; \exp\left[-\frac{(1-\Lambda^2)(x-\Lambda^n x_0)^2}{2\epsilon(1-\Lambda^{2n})}\right] .
\ee
Therefore, the trace (\ref{trace-int}) of the $n^{\rm th}$ iterate of the Frobenius-Perron operator is obtained as
\be
{\rm tr}\, \hat P^n = \int dx \, K_n(x,x) =  \frac{1}{\vert 1-\Lambda^n\vert} \; .
\label{trace-pt}
\ee
Two generic cases arise.  

If $\vert\Lambda\vert <1$, the fixed point $x=0$ is attracting and the Fredholm determinant (\ref{det-tr}) becomes
\be
\det\left(\hat I - \chi^{-1}\, \hat P\right)=\prod_{k=0}^{\infty} \left( 1 - \frac{\Lambda^k}{\chi}\right) = 0
\ee
so that the eigenvalues are given by
\be
\chi_k = \Lambda^k \qquad\mbox{with}\qquad k=0,1,2,...
\ee
The leading eigenvalue $\chi_0$ is equal to unity so that the probability density converges towards the Gaussian invariant density.

If $\vert\Lambda\vert >1$, the fixed point $x=0$ is unstable and the Fredholm determinant becomes
\be
\det\left(\hat I - \chi^{-1}\, \hat P\right)=\prod_{k=0}^{\infty} \left( 1 - \frac{1}{\chi}\, \frac{1}{\vert\Lambda\vert\Lambda^k}\right) = 0
\ee
so that the eigenvalues are now given by
\be
\chi_k = \frac{1}{\vert\Lambda\vert\Lambda^k} \qquad\mbox{with}\qquad k=0,1,2,...
\ee
Here, the leading eigenvalue $\chi_0$ is strictly less than unity so that the process is nonstationary with the positive  escape rate $\gamma=-\ln\chi_0=\ln\vert\Lambda\vert$.

The symplectic map (\ref{sympl_map}) ruling the classical orbits in the weak-noise limit takes the form
\be
\left(
\begin{array}{c}
x_{i+1}\\
p_{i+1}
\end{array}
\right)
=
\left(
\begin{array}{cc}
\Lambda & \Lambda^{-1}\\
0 & \Lambda^{-1}
\end{array}
\right)
\left(
\begin{array}{c}
x_{i}\\
p_{i}
\end{array}
\right) .
\ee
This two-dimensional map has a unique fixed point at the origin $x=p=0$.
The direction $p=0$ corresponds to the eigenvalue $\Lambda$ and
the direction $p = (1-\Lambda^2)\, x$ to the eigenvalue $\Lambda^{-1}$.

\section{Escape rate in the continuous limit}
\label{AppB}

For both the noisy logistic map (\ref{logistic_map}) and the exponential map (\ref{exp_map}), the two fixed points $a$ and $b$ coincide at a critical value of the control parameter $\mu$.  As shown in Ref.~\cite{DG09}, the noisy map (\ref{noisy_map}) behaves in this limit as the Langevin process ruled by the stochastic differential equation
\be
\frac{dx}{dt} = g(x) + \eta(t) \qquad\mbox{with} \qquad g(x)=f(x)-x
\ee
 where $t$ is a continuous time interpolating between its integer values $t=n$ at which $x(t)=x_n$
 and $\eta(t)$ is a Gaussian white noise such that 
 \be
 \langle\eta(t)\rangle=0 \qquad\mbox{and}\qquad  \langle\eta(t)\,\eta(t')\rangle=\epsilon \, \delta(t-t') \, .
 \ee

The potential
\be
U(x) = - \int g(x) \, dx
\ee
is introduced such that
\bea
&& U'(x) = -g(x) = x-f(x) \, ,\\
&& U''(x) = -g'(x) = 1-f'(x) \, .
\eea
If this potential has its minimum at $a$ and its maximum at $b$, the escape rate is given by \cite{C43}
\be
\gamma = \frac{1}{2\pi} \, \sqrt{ U''(a) \, \vert U''(b)\vert} \, \exp\left\{ -\frac{2}{\epsilon}\left[ U(b)-U(a)\right]\right\} .
\ee
In terms of the function $f(x)$ defining the map, we get
\be
\gamma = \frac{1}{2\pi} \, \sqrt{ \left[1-f'(a)\right]\, \vert 1-f'(b)\vert} \, \exp\left\{ \frac{2}{\epsilon}\int_{a}^{b} \left[ f(x)-x\right]\right\}
\ee
where the minimum and the maximum are the fixed points of the noiseless map, respectively, $a=f(a)$ and $b=f(b)$.

\subsection{Noisy logistic map}

Let us take the logistic map (\ref{logistic_map}).  The corresponding potential is given by
\bea
&& U(x) = \frac{1-\mu}{2}\, x^2 + \frac{\mu}{3} \, x^3 \, ,\\
&& U'(x) = (1-\mu)\, x + \mu \, x^2 \, ,\\
&& U''(x) = 1-\mu + 2 \, \mu \, x \, .
\eea
For $\mu>1$, the stable and unstable fixed points are respectively:
\bea 
&& a = 1-1/\mu \, ,\qquad U(a) = \frac{(\mu-1)^3}{6\,\mu^2} \, ,\qquad U''(a) = \mu-1 \, ,\\
&& b = 0\, , \qquad\qquad\quad  U(b) = 0 \, ,\qquad\qquad\quad U''(b) = 1-\mu \, ,
\eea
and vice versa for $\mu<1$.
Therefore, the escape rate is given by
\be
\gamma = \frac{\vert\mu-1\vert}{2\pi} \, \exp\left(- \frac{\vert\mu-1\vert^3}{3\, \epsilon\, \mu^2}\right)
\label{esc_rate-logistic_map}
\ee
near $\mu=1$.

\subsection{Noisy exponential map}

Let us take the exponential map (\ref{exp_map}).  The corresponding potential is given by
\bea
&& U(x) = \frac{1}{2}\, x^2 + \mu \, (x+1) \, \exp(-x) \, ,\\
&& U'(x) = x-\mu \, x \, \exp(-x) \, ,\\
&& U''(x) = 1+\mu\, (x-1)\, \exp(-x) \, .
\eea
For $\mu>1$, the stable and unstable fixed points are respectively:
\bea 
&& a = \ln\mu \, ,\qquad U(a) = 1+\ln\mu+\frac{1}{2}\, (\ln\mu)^2  ,\qquad U''(a) = \ln\mu \, ,\\
&& b = 0 \, , \qquad\quad  U(b) = \mu \, ,\qquad\qquad\qquad\qquad\qquad U''(b) = 1-\mu \, ,
\eea
and vice versa for $\mu<1$.
Therefore, the escape rate is given by
\be
\gamma = \frac{1}{2\pi} \, \sqrt{\vert\mu-1\vert\, \vert\ln\mu\vert}\, \exp\left[- \frac{1}{\epsilon}\Big\vert 2\mu - 2 -2\ln\mu-(\ln\mu)^2\Big\vert\right]
\label{esc_rate-exp_map}
\ee
near $\mu=1$.


\end{document}